\documentclass[twocolumn,showpacs,preprintnumbers,amsmath,amssymb]{revtex4-1} \usepackage{dcolumn}

\usepackage{graphicx} \usepackage{bm}

\begin{document}

\title{On calculation of RKKY range function in one dimension}
\date{\today}
\author{Tomasz M. Rusin$^1$} \author{Wlodek Zawadzki$^2$}

\email{Tomasz.Rusin@orange.com}
\affiliation{$^1$ Orange Poland sp. z o. o., Al. Jerozolimskie 160, 02-326 Warsaw, Poland \\
             $^2$ Institute of Physics, Polish Academy of Sciences, Al. Lotnik\'ow 32/46, 02-688 Warsaw, Poland}


\begin{abstract}
The effect of strong singularity in
the calculation of range function for the RKKY interaction in 1D electron gas is discussed.
The method of handling this singularity is presented.
A possible way of avoiding the singularity in the Ruderman-Kittel perturbation theory in 1D is described.
\end{abstract}

\maketitle

\section{Introduction}
Some years after the discovery of Ruderman-Kittel-Kasuya-Yosida (RKKY)
interaction between localized magnetic moments in three dimensions~\cite{Ruderman1954},
Kittel considered an extension of this interaction to lower dimensional system~\cite{Kittel1968}.
In the late 1980's and beginning of the 1990's
the RKKY interaction was recognized as one of the mechanisms of coupling between magnetic layers in metallic
superlattices~\cite{Bruno1992}, and the energy of RKKY interaction in quasi 1D systems
was determined experimentally by Parkin and Mauri~\cite{Parkin1991}.
A review of these efforts is summarized in Ref.~\cite{Yafet1994}.
Later, the RKKY interaction in 1D or quasi-1D systems was investigated
in many other works, see e.g.~\cite{Imamura2004},
and this subject is of actuality until present days, see e.g.~\cite{Nejati2017}.
For this reason, all subtleties of this problem should be clarified.

In his work, Kittel calculated the energy of RKKY interaction in one dimension  between
two localized magnetic moments embedded in a free electron gas~\cite{Kittel1968}.
He calculated first the
magnetic susceptibility~$\chi(q)$ of the electron gas in the presence of magnetic moments
and then the range function was obtained as the Fourier transform of~$\chi(q)$.
In the appearing integral Kittel changed the order of integration which lead
to erroneous results predicting a finite interaction energy
at infinite distance between localized moments. This error was corrected
in the Erratum to Ref.~\cite{Kittel1968}, and the correct result was obtained with a reverse
order of integration. Some time later Yafet~\cite{Yafet1987} showed that the problem
reported by Kittel is caused by the presence of a strong
singularity of the double integral at~$k=q=0$
and, because of the singularity, it is not allowed to change the order
of integration over~$k$ and~$q$ variables.
To show this, Yafet calculated twice the range function taking different orders of
integrations and obtained different results.
Then he determined the correct order of integrations.
Further subtleties of this problem were
discussed by Guliani {\it et al.}~\cite{Giuliani2005}.
Litvinov and Dugaev~\cite{Litvinov1998} showed that an application of
Green's function formalism allows one to avoid singularities at~$k=q=0$.

There exists an alternative method to calculate the RKKY interaction
proposed in the original approach of Ruderman and Kittel (RK)
to the 3D case~\cite{Ruderman1954}.
This method is based on a direct calculation of the second order correction to the energy of
free electron gas in the presence of two localized magnetic moments.
In 3D one obtains a double integral over~$|k'|> k_F$ and~$|k|\leq k_F$ domain,
which does not contain the strong singularity. This integral
is then replaced by a difference of two integrals. Applying this procedure
to 1D gas one finds that, surprisingly,
each of the two integrals contains a strong singularity at~$k=k'=0$.
This singularity does not exists in 2D or 3D cases.
But in the 1D case there appears a singularity which is analogous to that
appearing in the calculation of the range function in one dimension
with the use of susceptibility~$\chi(q)$ discussed by Yafet~\cite{Yafet1987}.

In the present note we analyze the effect of strong singularity at~$k=k'=0$
on the range function of the RKKY interaction in 1D calculated with the use of~RK approach.
Our results extend previous analyzes of singularities appearing in
the calculations of the range function with use of susceptibility~$\chi(q)$ in 1D,
as described in Refs~\cite{Kittel1968,Yafet1987,Giuliani2005}.
Then we show the effect of the order of integration over the singular part of the integral
in the 1D case and determine the correct order of integration.
Finally we propose another way to calculate the range function using a domain that is free of
strong singularities.

\section{Theory}
Let us consider a one-dimensional free electron gas.
Let the two spins~$\hat{\bm S}_i$ be located at~${\bm R}_i$, where~$i=1,2$.
A coupling between the conduction electrons and the localized spins is assumed in the form of
s-d interaction
\begin{equation}
 \hat{H}_{sd} = \frac{J_{sd}}{N_{1D}} \sum_{i=1,2}
  \delta({\bm R} - {\bm R}_i)\hat{\bm S}_i\hat{\bm \sigma},
\end{equation}
where~$\hat{\bm \sigma}$ is electron spin operator,~$J_{sd}$ is the energy of s-d coupling,
and~$N_{1D}$~is the one-dimensional density of magnetic atoms. Note that~$J_{sd}/N_{1D}$ has the
dimensionality of~[energy]~$\times$ [length].
Following Ruderman and Kittel, the second order correction to the energy of electron
gas perturbed by localized spins is~\cite{Ruderman1954}
\begin{equation} \label{DeltaE}
 \Delta E^{(2)} = \frac{J_{sd}^2}{(2\pi)^2 N_{1D}^2}\frac{2m^*}{\hbar^2} \hat{\bm S}_i\hat{\bm S}_j F_{1D}(r)
\end{equation}
where
\begin{equation} \label{Fr0}
 F_{1D}(r) = \int_{-k_F}^{k_F} \hspace{-1em} dk \left[ \left( \int_{-\infty}^{-k_F} \hspace{-1.5em} +
 \int_{k_F}^{\infty} \right)
 \frac{\cos(kr)\cos(k'r)}{k'^2-k^2} dk' \right], \ \
\end{equation}
in which~$m^*$ is the electron effective mass,~$k_F$ is the Fermi vector,~$r=R_i-R_j$, and~$F_{1D}(r)$
is the so-called range function. The order of integration in Eq.~(\ref{Fr0}) follows from
the method of calculation of~$\Delta E^{(2)}$:
first one selects the wave vector~$k$, calculates the second order correction~$\Delta E^{(2)}_k$
to the electron's energy~$E_k$ [square bracket in Eq.~(\ref{Fr0})],
and then sums~$\Delta E^{(2)}_k$ over~$k$ within the 1D Fermi sphere.
Considering Eq.~(\ref{Fr0}) one concludes that,
since the~$k$ vectors are inside the 1D Fermi sphere and the~$k'$ vectors are
outside the sphere, the denominators in Eq.~(\ref{Fr0}) are always nonzero
and no singularity occurs.

\begin{figure} \includegraphics[scale=0.30]{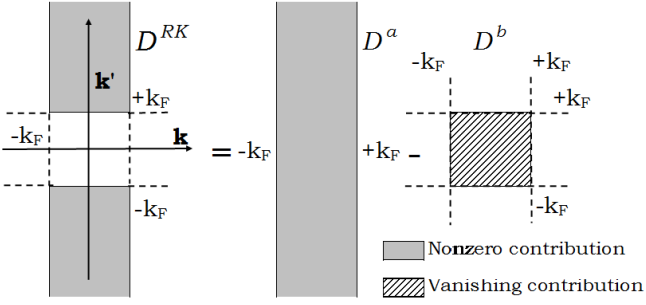}
 \caption{Schematic visualization of integration domain defined in~(\ref{DIRK})--(\ref{DIb}).
         Left side of equation: domain of integration
         in Eq.~(\ref{Fr0}) (grey), right side: two domains of
         integration proposed in Ref.~\cite{Ruderman1954}, gray and dotted.
         Grey areas give nonzero contribution to the range function while
         integral over dotted areas vanishes due to symmetry.} \label{Fig1}
\end{figure}

The difficulty in Eq.~(\ref{Fr0}) is that the integral over~$dk'$
can not be calculated analytically. To overcome this problem RK~\cite{Ruderman1954}
proposed to replace the integral in Eq.~(\ref{Fr0}) over the domain
\begin{equation} \label{DIRK}
 {\cal D}^{RK}: (k,k') \in [-k_F,k_F] \times \mathbb{R} \setminus [-k_F,k_F],
\end{equation}
by the difference of two integrals over domains
\begin{eqnarray}
 {\cal D}^a &:& (k,k') \in [-k_F, k_F] \times \mathbb{R},  \label{DIa} \\
 {\cal D}^b &:& (k,k') \in [-k_F, k_F] \times [-k_F, k_F], \label{DIb}
\end{eqnarray}
see Figure~\ref{Fig1}.
In the above expressions we used the notation of the set theory. As an example,
if~$k$ is a member of set~${\cal A}$,
the notation~$k \in {\cal A}$ is used. Similarly,~$\times$
denotes the cartesian product of two sets,~${\cal A}\setminus {\cal B}$
denotes difference between the two sets, and~${\cal A}\cup {\cal B}$
means the union of the two sets. For more detailed
description of set notion see Ref.~\cite{WikiSets}.

From~(\ref{DIRK})--(\ref{DIb}) we have
\begin{equation} \label{2Int}
F_{1D}(r) = \int {\cal D}^{RK} = \int {\cal D}^a - \int {\cal D}^b,
\end{equation}
in which we use the notation
\begin{equation} \label{Fr1}
 \int {\cal D}^a = \iint_{(k,k') \in {\cal D}^a}
 \left(\frac{\cos(kr)\cos(k'r)}{k'^2-k^2} dk'\right) dk,
\end{equation}
and similarly for~${\cal D}^b$ and~${\cal D}^{RK}$.
This method works correctly for 3D. However, doing so for 1D
requires caution due to the presence of strong
singularity at~$k=k'=0$ in Eq.~(\ref{Fr1})
for the domains~${\cal D}^a$ and~${\cal D}^b$.
We show below that this method may not be directly applied to the 1D case
since the singularity at~$k=k'=0$ gives a nonzero contribution to the integrals.

Consider first~$\int {\cal D}^a$, as given in Eqs.~(\ref{DIa}) and (\ref{Fr1}).
The integral over~$k'$ is obtained with the use of formula~3.723.9 in~\cite{GradshteinBook}
\begin{equation}
 \label{Int3.723.9}
 \int_{-\infty}^{\infty} \frac{\cos(rk')}{k^2-k'^2}dk' = \frac{\pi}{k}\sin(rk),
\end{equation}
which is valid for~$|r|, |k| > 0$. Then
\begin{equation} \label{Ia}
 \int {\cal D}^a = -\pi \int_{-k_F}^{k_F} \frac{\cos(kr)\sin(kr)}{k} dk = -\pi {\rm Si}(2k_Fr),
\end{equation}
where~${\rm Si}(x) = \int_0^x (\sin(t)/t)dt$ is the sine-integral in the standard
notation, see~\cite{GradshteinBook}.

The subtle point in the derivation of Eq.~(\ref{Ia}) is that the integral on
the left hand side of Eq.~(\ref{Int3.723.9})
does not exist at~$k=0$, since for~$k=0$ and~$|k'|\rightarrow 0$ the
integrand diverges as~$1/k^{'2}$.
Therefore Eq.~(\ref{Int3.723.9}) in valid for all~${\cal D}^a$ except in the small domain
\begin{equation}
 {\cal D}^{\epsilon}:(k,k') \in [-\epsilon, \epsilon] \times [-\epsilon, \epsilon], \label{DIe}
\end{equation}
with~$\epsilon \rightarrow 0$, for which the identity~(\ref{Int3.723.9})
can not be used. To overcome this problem we isolate the domain~${\cal D}^{\epsilon}$
out of the integration
domain:~$\int {\cal D}^a = \int {\cal D}^{a\epsilon} + \int {\cal D}^{\epsilon}$,
in which:~${\cal D}^{a\epsilon} = {\cal D}^a \setminus {\cal D}^{\epsilon}$.
The contribution to the range function coming from~${\cal D}^{\epsilon}$
has to be calculated separately.

\begin{figure} \includegraphics[scale=0.29]{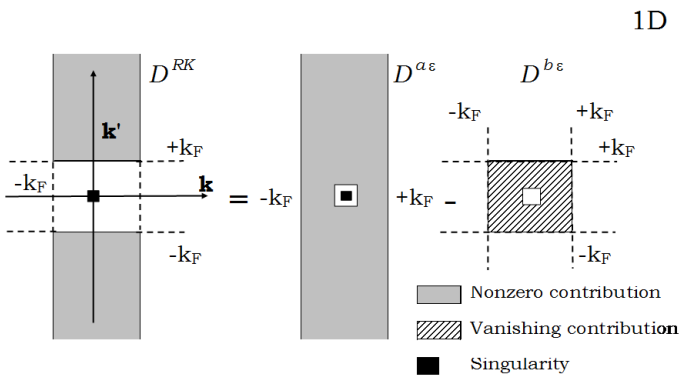}
 \caption{Schematic visualization of difference of the two domains shown in Eq.~(\ref{4Int}).
          Grey and dotted areas have the same meaning as in Figure~1.
          Black squares: strong singularity at~$k=k'=0$. Note that the two domains
          on the rhs still do not include strong singularity.}~\label{Fig2}
\end{figure}

\begin{figure} \includegraphics[scale=0.26]{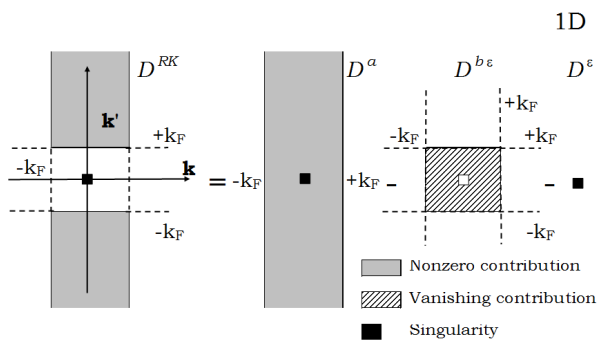}
 \caption{Schematic visualization of difference of the three domains shown in~(\ref{3Int}).
          Grey and dotted areas have the same meaning as in Figure~\ref{Fig1}.
          Note that on the rhs the strong singularity (black square)
          is added to~${\cal D}^{a}$
          and explicitly subtracted. This is the main difference between the 3D case
          in which the strong singularity does not exist, see text.}~\label{Fig3}
\end{figure}

Turning to~$\int {\cal D}^b$ we note that there is a similar problem
with the singularity at~$k=k'=0$,
so that we again isolate~${\cal D}^{\epsilon}$ out of the integration
domain:~$\int {\cal D}^b = \int {\cal D}^{b\epsilon} + \int {\cal D}^{\epsilon}$
in which:~${\cal D}^{b\epsilon} = {\cal D}^b \setminus {\cal D}^{\epsilon}$.
Let us assume that the integral~${\cal D}^{\epsilon}$ is finite, which is crucial for
the calculations. Then from Eq.~(\ref{2Int}) we have (see Figure~\ref{Fig2})
\begin{eqnarray} \label{4Int}
 F_{1D}(r) &=& \int \left( {\cal D}^{a\epsilon} \cup {\cal D}^{\epsilon} \right)
        - \int \left( {\cal D}^{b\epsilon} \cup {\cal D}^{\epsilon} \right) \nonumber \\
        &=& \int {\cal D}^{a\epsilon} - \int {\cal D}^{b\epsilon}.
\end{eqnarray}
Thus, if the integral~$\int {\cal D}^{\epsilon}$ is finite,
the contribution arising from the two integrals~$\int {\cal D}^{\epsilon}$
in Eq.~(\ref{4Int}) cancels out. However, in order to apply Eq.~(\ref{Ia})
one has to calculate the integral over the domain~${\cal D}^a$
instead of~${\cal D}^{a\epsilon}$. Assuming that~$\int {\cal D}^{\epsilon}$ is
finite we can rewrite Eq.~(\ref{4Int}) as
\begin{equation} \label{3Int}
 F_{1D}(r) = \int {\cal D}^a - \int {\cal D}^{b\epsilon} - \int {\cal D}^{\epsilon},
\end{equation}
in which~$\int {\cal D}^a$ is given in Eq.~(\ref{Ia}), see Figure~\ref{Fig3}.
This is the final result of our manipulations. Since there is no strong
singularity in~${\cal D}^{b\epsilon}$, we may apply the method proposed in Ref.~\cite{Ruderman1954}
and show that~$\int {\cal D}^{b\epsilon}=0$, see Appendix~A.
Comparing Eq.~(\ref{3Int}) with Eq.~(\ref{DIRK}) we note the additional contribution
in 1D from~$\int {\cal D}^{\epsilon}$ to the range function, which does not exist in 3D,
see Figures~\ref{Fig1} and~\ref{Fig3}.

To calculate~$\int {\cal D}^{\epsilon}$ we use a similar approach
to that applied by Yafet~\cite{Yafet1987}.
We first approximate in Eq.~(\ref{Fr1}):~$\cos(kr) \simeq 1$ and~$\cos(k'r) \simeq 1$,
which is valid for sufficiently small~$|k|$ and~$|k'|$. Then we have
\begin{equation} \label{Ieps1}
 \int {\cal D}^{\epsilon} =\int_{-\epsilon}^{\epsilon} \hspace{-0.5em} dk
 \int_{-\epsilon}^{\epsilon} \frac{dk'}{k'^2-k^2}.
\end{equation}
Using the identity:~$1/(k'^2-k^2)=(1/2k)[1/(k'-k)-1/(k'+k)]$
and integrating in Eq.~(\ref{Ieps1}) over~$k'$ we find
that~$\int {\cal D}^{\epsilon}$ is nonzero
\begin{eqnarray} \label{Ieps2}
 \int {\cal D}^{\epsilon}= \int_{-\epsilon}^{\epsilon}
 \left( \frac{\ln |\epsilon-k|}{k}       -\frac{\ln|\epsilon+k|}{k} \right) dk =\nonumber \\
 \int_{-1}^{1} \left(\frac{\ln |1-u|}{u} -\frac{\ln|1+u|}{u} \right) du =\nonumber \\
 = -2{\rm Li}_2(1) + 2{\rm Li}_2(-1) = -\frac{\pi^2}{2}.
\end{eqnarray}
This is the peculiarity of 1D case, which does not appear in 2D and 3D, see Appendix~B.
In the above equation:~${\rm Li}_2(x)= -\int_{0}^{x}du \ln|1-u|/u$ is the dilogarithm function,
see~\cite{GradshteinBook,Mitchell1949,LewinBook}, and we have
used:~${\rm Li}_2(1) = \pi^2/6$ and:~${\rm Li}_2(-1)=-\pi^2/12$, see~\cite{GradshteinBook}.
Collecting the results from Eqs.~(\ref{Ia}),~(\ref{3Int}) and~(\ref{Ieps2}) we have
\begin{equation} \label{Fr2}
 F_{1D}(r) = \pi \left[\frac{\pi}{2} - {\rm Si}(2k_Fr) \right],
\end{equation}
which agrees with the results reported in the
literature~\cite{Yafet1987,Giuliani2005,Litvinov1998}.
The range function in Eq.~(\ref{Fr2}) oscillates with the period:~$T_r=\pi/k_F$
and decays to zero at large distances between spins.
Note that neglecting the contribution from~$\int {\cal D}^{\epsilon}$ one
erroneously obtains:~$F_{1D}(r)=\int {\cal D}^a \propto {\rm Si}(2k_Fr)$, see~\cite{Kittel1968},
which for large~$r$ tends to a finite value.

\begin{figure} \includegraphics[scale=0.19]{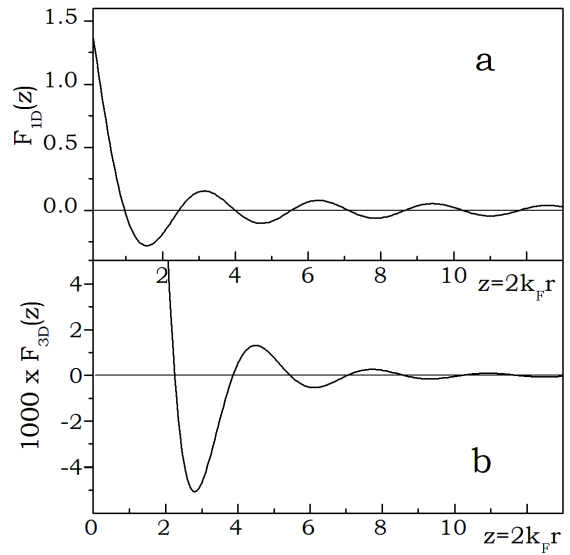}
 \caption{Upper panel: The range function~$F_{1D}(r)$, as given in Eq.~(\ref{Fr2}).
          Lower panel: The range function~$F_{3D}(r)$, see text.
          Please note the difference in scales between two panels}~\label{Fig4}
\end{figure}

In order to  illustrate~$F_{1D}(r)$ we plot this function in Figure~\ref{Fig4}a,
and compare it with the widely-known
range function in 3D:~$F_{3D}(z)=[\sin(z)-z\cos(z)]/z^4$ with~$z=2k_Fr$,
see Figure~\ref{Fig4}b.
As seen in the Figures, both functions have the same oscillation period,
and both vanish at~$k_Fr \rightarrow \infty$, but
the function~$F_{1D}(r)$ decays as~$r^{-1}$, i.e. much slower than~$F_{3D}(r)$.

In the calculation of~$\int {\cal D}^{\epsilon}$ it is not
allowed to change the order of integration over~$k$ and~$k'$ variables. To
show this we calculate an integral~$\int {\cal D}^{\epsilon}_R$
in analogy to that in Eq.~(\ref{Ieps1}), but with the reversed order of integration
over~$k$ and~$k'$. Using the identity:~$1/(k'^2-k^2)=(1/2k')[1/(k'-k)+1/(k'+k)]$ one obtains
\begin{eqnarray} \label{ISR}
 \int {\cal D}^{\epsilon}_R = \int_{-\epsilon}^{\epsilon}
 \left( \frac{-\ln |\epsilon-k'|}{k'} + \frac{\ln|\epsilon+k'|}{k'} \right) dk' =\nonumber \\
 = 2{\rm Li}_2(1) - 2{\rm Li}_2(-1) = +\frac{\pi^2}{2}.
\end{eqnarray}
Thus~$\int {\cal D}^{\epsilon}_R \neq \int {\cal D}^{\epsilon}$, so the change in
the order of integration over~$k$ and~$k'$ is not allowed.

\begin{figure} \includegraphics[scale=0.30]{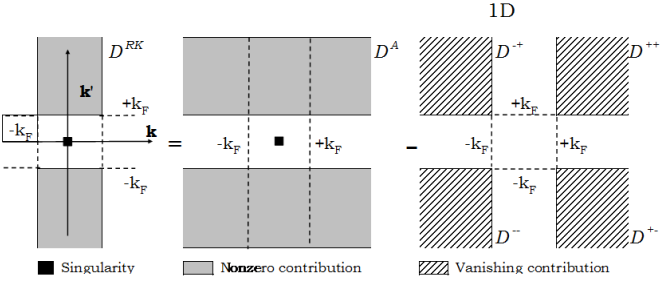}
 \caption{Schematic visualization of integration domain defined in~(\ref{5Int}).
          Grey and dotted areas have the same meaning as in Figure~\ref{Fig1}.
          Note that on the rhs the domains of integration do not include the strong
          singularity, so one can safely change the order of integration over~$k$ and~$k'$,
          see Eq.~(\ref{IA}).}~\label{Fig5}
\end{figure}

In 1D one can avoid the problem with the strong singularity at~$k=k'=0$
by replacing the domain~${\cal D}^{RK}$ in Eq.~(\ref{DIRK}) by another one,
still free of the strong singularity. For example, one can
choose domain~${\cal D}^f$ defined as
\begin{eqnarray}
 {\cal D}^f &:& {\cal D}^A \setminus \left( {\cal D}^{++} \cup {\cal D}^{+-}
        \cup {\cal D}^{-+} \cup {\cal D}^{} \right) \label{5Int}, \\
 {\cal D}^A  &:& (k,k') \in \mathbb{R} \times
               \left([k_F,\infty) \cup [k_F,\infty) \right), \label{DIA} \\
 {\cal D}^{++} &:& (k,k') \in [k_F,\infty) \times [k_F,\infty). \label{DI++}
\end{eqnarray}
The domain~${\cal D}^{++}$ describes the right upper corner of the~$k\times k'$ plane,
while domains~${\cal D}^{+-}$,~${\cal D}^{-+}$ and~${\cal D}^{--}$ describe
its three remaining corners, see Figure~\ref{Fig5}.
Since there are no strong singularities in any of the above domains,
in each domain of~(\ref{5Int})--(\ref{DI++}) it is allowed to change the order of integration
over~$k$ and~$k'$ vectors. Using similar arguments to those in Appendix~A
we obtain:~$\int ({\cal D}^{++} \cup {\cal D}^{+-} \cup {\cal D}^{-+} \cup {\cal D}^{--}) =0$,
and then:~$\int {\cal D}^f= \int {\cal D}^A$.
Changing the order of integration in~${\cal D}^A$ and calculating first the integral over~$k$
with use of Eq.~(\ref{Int3.723.9}) we find
\begin{equation} \label{IA}
 F_{1D}(r) = \pi\! \int_{k_F}^{\infty}\! \frac{\sin(2k'r)}{k'} dk' =
 \pi \left(\frac{\pi}{2} -{\rm Si}(2k_Fr)\right),
\end{equation}
in which~$\lim_{x\rightarrow \infty}{\rm Si}(x) =\pi/2$, see~\cite{GradshteinBook}.
This agrees with Eq.~(\ref{Fr2}). Note that for~${\cal D}^A$ there is
always:~$|k'|> k_F > 0$ and the integrand over~$k$ on the left hand side of Eq.~(\ref{Int3.723.9})
exists for all~$k'$ in the domain~${\cal D}^A$.

Comparing Figures~\ref{Fig1},~\ref{Fig2} and~\ref{Fig3} with Figure~\ref{Fig5}
we note that the transformed domains on the right-hand sides of
Figures~\ref{Fig1},~\ref{Fig2} and~\ref{Fig3} are 'vertical' in
the~$k-k'$ plane, while the corresponding domain in Figure~\ref{Fig5} is 'horizontal' one.
This seemingly minor change allows one to avoid {\it any} singularity appearing for small values
of {\it both}~$k$ and~$k'$ vectors. Turning to the initial domain of integration, as indicated on the
left-hand side of Figure~\ref{Fig5}, we see that this domain is limited to~$|k| \le k_F$
and~$|k'|> k_F$, i.e. it does not include strong singularity at~$k=k'=0$.
For sufficiently large~$k_F$ the existence or no-existence of the singularity at the origin
should not alter the integration over the RK domain. Thus the singularity
is only an artefact appearing in 1D case without an impact on the range function~$F_{1D}(r)$.
But in the arrangement proposed by RK, as seen in Figures~\ref{Fig1},~\ref{Fig2} and~\ref{Fig3},
one replaces the singularity-free domain by a
combination of domains including the singularity, which requires strict mathematical rigor in handling
the problem. In contrast, in the arrangement shown in Figure~\ref{Fig5} one transforms
the singularity-free domain by a combination of five singularity-free domains,
and the correct results are obtained in a straightforward way, see Eq.~(\ref{IA}).

\section{Discussion and summary}

The problem arising in the calculation of interaction energy with the
use of the perturbation expansion,
as expressed in Eqs.~(\ref{DeltaE}) and~(\ref{Fr0}), is to justify a truncation
of the expansion to the second order terms.
In general, the perturbation series is
convergent if there exists a 'small parameter'~$\alpha \simeq V(r)/(E_k-E_{k'}) \ll 1$.
Turning to Eq.~(\ref{Fr0}) we may suspect that, possibly, the perturbation expansion may not
converge for states~$k'$ lying close to the Fermi sphere~$k_F$ since in this case~$E_k' \simeq E_k$,
and the denominator in Eq.~(\ref{Fr0}) is small.

To analyze this effect quantitatively we calculate a contribution
of~$\delta E^{(2)}$ to the interaction energy~$\Delta E^{(2)}$
arising from states~$k'$ belonging to small slices close to
the Fermi level:~$k' \in (k_F, k_F + \delta k] \cup [-k_F-\delta k, -k_F)$,
with~$\delta k_F \ll k_F$.
We define the integration domain
\begin{equation}
 {\cal D}^{\delta k'}: (k,k') \in [-k_F,k_F] \times
     \left( (k_F, k_F + \delta k] \cup [-k_F-\delta k, -k_F) \right),
\end{equation}
and calculate the range function~$F_{1D}(r)$ on this domain. The calculations are analogous to
those in Eqs.~(\ref{5Int})--(\ref{IA}),
but with the integration over~$k'$ limited to~$\pm (k_F+\delta k)$
instead of~$\pm \infty$, respectively.
Then we obtain from Eq.~(\ref{IA})
\begin{equation} \label{deltaE}
 \delta E^{(2)} \propto \pi\! \int_{k_F}^{k_F+\delta k}\! \frac{\sin(2k'r)}{k'} dk' \simeq
     \pi \frac{\sin(2k_Fr)}{k_F} \delta k.
\end{equation}

The the Fermi vector~$k_F=2\pi/\lambda_F$ entering into the RKKY range
function in Eqs.~(\ref{IA}) and~(\ref{deltaE})
was first measured directly by Parkin and Mauri in Ni$_{80}$Co$_{20}$/Ru
superlattices~\cite{Parkin1991}.
The authors reported~$\lambda_F=11.5$~\AA, which gives~$k_F=0.55$\AA$^{-1}$.
Other values found in the literature are on
the order of~$k_F\simeq 0.5$~\AA$^{-1}$--$1.2$~\AA$^{-1}$,
see Ref.~\cite{Yafet1994} and references therein.
For such values of~$k_F$ the energy~$\delta E^{(2)}$ in Eq.~(\ref{deltaE})
does not diverge and the second order perturbation
approach is justified.

A contribution of third-order terms to RKKY in 3D was calculated in
 Ref.~\cite{Vertogen1966} and it turned out that
these terms are divergent at the limit~$k'\rightarrow \infty$ of integration
over excited states~$k'$. This may possibly occur
also in 1D case. However, as shown in~\cite{Bowen1968},
the motion of atoms due to phonons removes
the divergence in the third-order energy. On the other hand,
an approximation of the realistic energy
bands by the parabolic dispersion is valid only up to a certain value of~$k_{max}$,
which may not exceed
edges of the Brillouin zone:~$k_{BZ} \simeq 2\pi/a_{latt} \simeq 2.5$~\AA$^{-1}$
for typical values of lattice
constants~$a_{latt}$. Therefore, the divergence appearing
for~$k'\rightarrow \infty$ is not physical.
Introducing a reasonable cut-off in the~$k'$ integration,
or taking more realistic (e.g. tight-binding like) energy
dispersion, one obtains finite results for all dimensions.
Thus introducing the cut-off in the calculation of
third-order terms, a strong singularity at~$k=k'=0$ may also be removed
by methods discussed in our paper. The resulting interaction
would include higher powers of~$\hat{\bm S}_1\hat{\bm S}_2$
operators, see e.g.~\cite{Doman1969}.

In summary, we analyzed the effect of strong singularity in the calculation of
range function for RKKY interaction in one dimension using the Ruderman-Kittel method.
This approach is complementary to the more frequently used method based
on the susceptibility of the free electron gas.
It is pointed out that, in the RK method applied to the one-dimensional gas,
the initial singularity-free integral
is replaced by two integrals, each of them including strong singularity at~$k=k'=0$.
The way of isolating the singular parts of the two integrals is derived and
the method of handling the singularity is described. It is shown that the integral over
the singularity depends on the order of integration over~$k$ and~$k'$
vectors and the correct order of integration is determined.
The reason for disappearance of the singularity in higher dimensions
is explained. Importantly, a possible way of avoiding the singularity in one dimension
is proposed, see Figure~\ref{Fig5}.
Our analysis should help to avoid similar difficulties
which may occur in other low-dimensional systems.

\appendix
\section{}

We show that~$\int {\cal D}^{b\epsilon}=0$, see Eq.~(\ref{Ieps2}).
Let~$J^{b\epsilon}_{k'k} = \int {\cal D}^{b\epsilon}$,
where the lower indices define the order of
calculation in the integrals. By changing variables:~$(k,k')\rightarrow (k',k)$
we find:~$J^{b\epsilon}_{k'k}=-J^{b\epsilon}_{kk'}$, because of the change of
signs in the denominators, see Eq.~(\ref{Ieps2}).
Since there is no strong singularity in~${\cal D}^{b\epsilon}$,
the integral~$\int {\cal D}^{b\epsilon}$ does not depend on the order of
integration over~$k$ and~$k'$ variables.
Then we have:~$J^{b\epsilon}_{k'k}=J^{b\epsilon}_{kk'}$ which
gives the desired
result:~$J^{b\epsilon}_{kk'}=-J^{b\epsilon}_{kk'} \Leftrightarrow \int {\cal D}^{b\epsilon}=0$.
This also occurs for integrals over any domain~${\cal D}^s$ symmetric within~$k$ and~$k'$
variables. Using the same arguments one may show that~$\int {\cal D}^{b\epsilon}=0$
and~$\int {\cal D}^{++} + {\cal D}^{+-} + {\cal D}^{-+} +{\cal D}^{--} =0$, see Eq.~(\ref{DI++})
and Figure~\ref{Fig5}.

\section{}

The problem with the integration over~${\cal D}^{\epsilon}$ does not exist
in two and three dimensions
since in these cases there is no strong singularity at~$k=k'=0$. In this Appendix we
quote for completeness the corresponding calculations.
In 3D, after integration over angular variables,
one obtains for the range function (see Eq.~(6) in~\cite{Ruderman1954}),
\begin{equation} \label{F3D1}
 F_{3D}(r) = \int_{-k_F}^{k_F} \int_{-k_F}^{k_F} \frac{e^{ir(k+k')}kk'}{k^{'2}-k^2} dk' dk,
\end{equation}
which has no contribution from the singularity at~$k=k'=0$ because of the $kk'$ factor in the integrand.
To show this we calculate the integral in Eq.~(\ref{F3D1}) over
domain~${\cal D}^{\epsilon}$, see Eq.~(\ref{Ieps1}).
For small~$|k|$ and~$|k'|$ there is:~$e^{ir(k+k')} \rightarrow 1$ and,
instead of Eqs.~(\ref{Ieps1})--(\ref{Ieps2}), we have
\begin{equation} \label{I3D}
 \int {\cal D}^{\epsilon}_{3D} = \int_{-\epsilon}^{\epsilon} \int_{-\epsilon}^{\epsilon}
 \frac{kk'}{k^{'2}-k^2} dk' dk = 0,
\end{equation}
Thus, there indeed is no contribution from
the singularity at~$k=k'=0$. The same result is obtained for the reversed order
of calculation in the integrals in Eq.~(\ref{I3D}), so that~$\int {\cal D}^{\epsilon}_{3D}$
does not depend on the order of integration over~$k$ and~$k'$.
Similar arguments can be used for calculating the range function in 2D, in which
also the volume element~$kk'\ dk dk'$ appears.

\end{document}